\documentclass[twocolumn,showpacs,preprintnumbers,amsmath,amssymb,prl]{revtex4}

\usepackage{graphicx}
\usepackage{dcolumn}
\usepackage{bm}

\begin{document}

\title{Narrowing of the Balance Function with Centrality in Au+Au Collisions
at $\sqrt{s_{NN}}$ = 130 GeV}

\author{
J.~Adams$^3$, C.~Adler$^{11}$, Z.~Ahammed$^{23}$, C.~Allgower$^{12}$, 
J.~Amonett$^{14}$,
B.D.~Anderson$^{14}$, M.~Anderson$^5$, G.S.~Averichev$^{9}$, 
J.~Balewski$^{12}$, O.~Barannikova$^{9,23}$
% ...
}
\noaffiliation{}
\author{STAR Collaboration}
\affiliation{
% ...
$^3$University of Birmingham, Birmingham, United Kingdom\\
% ...
$^5$University of California, Davis, California 95616\\
% ...
$^{9}$Laboratory for High Energy (JINR), Dubna, Russia\\
% ...
$^{11}$University of Frankfurt, Frankfurt, Germany\\
$^{12}$Indiana University, Bloomington, Indiana 47408\\
% ...
$^{14}$Kent State University, Kent, Ohio 44242\\
% ...
$^{23}$Purdue University, West Lafayette, Indiana 47907\\
% ...
}

\date{\today}

\begin{abstract}
The balance function is a new observable based on the principle that
charge is locally conserved when particles are pair produced. Balance
functions have been measured for charged particle pairs and identified
charged pion pairs in Au+Au collisions at $\sqrt{s_{NN}}$ = 130 GeV at
the Relativistic Heavy Ion Collider using STAR.  Balance functions for
peripheral collisions have widths consistent with model predictions
based on a superposition of nucleon-nucleon scattering.  Widths in
central collisions are smaller, consistent with trends predicted by
models incorporating late hadronization.
\end{abstract}


\maketitle

Collisions of Au nuclei at ultra-relativistic energies can produce large
energy densities and high temperatures. At these densities and
temperatures, the produced matter may be best portrayed by partonic
(quark/gluon) degrees of freedom as opposed to those characterizing a
hot hadronic phase.  This partonic phase would necessarily be followed
by a transition to normal hadronic particles, which are ultimately
measured \cite{QGP}. Numerous probes of the high density medium have
been proposed \cite{signatures}, including observables related to
fluctuations and correlations
\cite{charge_fluct,charge_fluct2,fluct3,fluct4,fluct5}.

One such observable, the balance function, is sensitive to whether the
transition to a hadronic phase was delayed, as expected if the
quark-gluon phase were to persist for a substantial time
\cite{balance_paper}.  The authors of Ref.  \cite{balance_paper}
formulate the balance function as follows in this paragraph.
As a result
of local charge conservation, when particles and their anti-particles
are pair produced, they are correlated initially in coordinate space. 
If hadronization occurs early, the members of a charge/anti-charge pair
would be expected to separate in rapidity due to expansion and
re-scattering in the strongly interacting medium.  Alternatively,
delayed hadronization would lead to a stronger correlation in rapidity
between the particles of charge/anti-charge pairs in the final state.
Measuring this correlation involves subtracting uncorrelated
charge/anti-charge pairs on an event-by-event basis.  The remaining
charge/anti-charge particle pairs are examined to determine the
correlation as a function of the relative rapidity, $\Delta y$, between
the members of the pairs.  The balance function is expected to be
narrower for a scenario with delayed hadronization, and is therefore
sensitive to the conjecture that a quark-gluon plasma may be
produced.

The balance function is a new observable for heavy ion
collisions. This function is similar in form to the observable used in
\cite{Drijard} to study the associated charged density balance from both
p+p and $e^{+}+e^{-}$ collisions \cite{Aihara}.

In this analysis the balance function is used to examine the pseudorapidity 
correlation of non-identified charged particles and the rapidity correlation 
of identified charged pions.  The balance function is calculated as:
\begin{eqnarray}
B = \frac{1}{2} \left\{ \frac{\Delta_{+-} - \Delta_{++}}{N_{+}}+\frac{\Delta_{-+} - \Delta_{--}}{N_{-}}\right\}
\end{eqnarray}
where $\Delta_{+-}$ in the case of identified charged pion pairs denotes
the number of identified charged pion pairs in a given rapidity range
$\Delta y = | y(\pi^{+}) - y(\pi^{-}) |$, similarly for $\Delta_{++}$,
$\Delta_{--}$, and $\Delta_{-+}$. In the case of non-identified charged
particle pairs, pseudorapidity ($\eta$) is used.  The terms
$\Delta_{+-}$, $\Delta_{++}$, $\Delta_{--}$, and $\Delta_{-+}$ are
calculated using pairs from a given event and the resulting
distributions are summed over all events.  Specifically, the
distribution $\Delta_{+-}$ is calculated by taking in turn each positive
particle in an event and incrementing a histogram in $\Delta y$
($\Delta\eta$) with respect to all the negative particles in that event.
 The distribution $\Delta_{+-}$ is then summed over all events.  A
similar procedure is followed for $\Delta_{++}$, $\Delta_{--}$, and
$\Delta_{-+}$.
$N_{+(-)}$ is the number of positive(negative) pions or
non-identified charged particles summed over all events.
The balance
function is calculated for all charged pion or charged particle pairs
for a given centrality bin.  The balance function, in addition to being
proportional to opposite charge correlations, is also proportional to
the acceptance for each $\Delta y$ or $\Delta\eta$ bin.  Thus the
measured balance functions presented in this paper are narrower than
model calculations assuming no detector acceptance effects.

The data used in this analysis were measured using the Solenoidal
Tracker at RHIC (STAR) detector for Au+Au collisions at $\sqrt{s_{NN}}$
= 130 GeV. 
The main detector used for this analysis was the Time
Projection Chamber (TPC) \cite{STARcdr} located in a solenoidal magnetic
field of 0.25 T \cite{TPCref}.  The TPC provided tracking information
for charged particles having transverse momentum $p_{t}  >$ 100 MeV/c
and $|\eta| < 1.3$ with complete azimuthal acceptance.  Minimum bias
events were defined by the coincidence of two Zero Degree Calorimeters
(ZDCs) \cite{ZDC} located $\pm$ 18 m from the center of the interaction
region.  Central collisions were defined using both the ZDCs and the
Central Trigger Barrel (CTB), an array of scintillator slats surrounding
the outside of the TPC.  A coincidence in the ZDCs accompanied by a
signal in the CTB above a threshold empirically determined to correspond
to approximately the 15\% most central events was used to define the
central collision trigger. Centrality was determined using the charged
particle multiplicity distributions.  Central collisions correspond to 0
- 10\%, mid-central 10 - 40\%, mid-peripheral 40 - 70\%, and peripheral
70 - 96\% of the integrated total.  Only events with five or more tracks
in the TPC were used in this analysis.

For this analysis 150,777 minimum bias and 111,142 centrally
triggered events were used. Only charged particle tracks having more
than 15 space points along the trajectory were accepted.  In addition
the reconstructed trajectory
was required to point within 1 cm of the
primary vertex. To suppress double counting due to track splitting, the
ratio of reconstructed space points to possible space points along the
track was required to be greater than 0.52. Tracks were further required
to have a momentum 0.1 $< p <$ 2 GeV/c. Simulations of TPC performance
indicate the tracking efficiency for tracks within this acceptance and
momentum range is 85-90\% \cite{calderon}.

Charged pion identification was accomplished by selecting particles
within 2 standard deviations of the expected $dE/dx$ for a given
momentum, which provided pion identification for momenta less than 0.7
GeV/c. Kaon contamination of the pions is negligible at low momentum and
is estimated to be 5\% at 0.7 GeV/c.  Electrons were removed by
requiring the measured $dE/dx$ to be more than 2 standard deviations
away from the expected value for an electron of the measured momentum.
An estimated 1\% of the electron contamination remains in the pion
sample after these cuts are made.  The lowest $\Delta\eta$ ($\Delta y$)
bin, which was estimated to contain 90\% of the remaining electron
contamination, was not used in the calculation of the width of the
balance function.

Theoretical simulations of the balance function for Au+Au collisions at
$\sqrt{s_{NN}}$ = 130 GeV were carried out using version 1.36 of HIJING
\cite{hijing} with the default settings for impact parameters ranging
from 0 to 15 fm.  A realistic distribution was used for event vertices
within the interaction region. HIJING  events were processed using GEANT
\cite{geant} and the TPC track reconstruction software.  The same cuts
were applied to the HIJING events that were applied to the data.  None
of the features of the balance functions computed from the filtered
simulations show any dependence on centrality \cite{marguerite}. Thus
the STAR results are compared to HIJING integrated over centrality.
\begin{figure}
\includegraphics[width=3in]{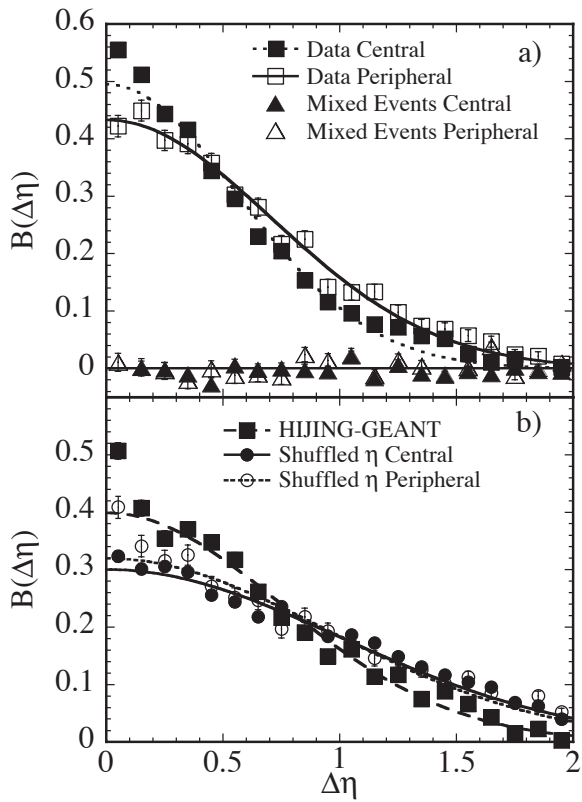}
\caption{\label{fig:balch} The balance function versus $\Delta\eta$ for
charged particle pairs from a) central and peripheral Au+Au collisions
at $\sqrt{s_{NN}}$ = 130 GeV and mixed events from central and
peripheral Au+Au collisions, and b) HIJING events filtered with GEANT
\cite{geant} and shuffled pseudorapidity events from central and
peripheral Au+Au collisions. To guide the eye, Gaussian fits excluding
the lowest bin in $\Delta \eta$ are shown. The error bars shown are
statistical. The balance function for HIJING events is independent of
centrality.}
\end{figure}
Figure \ref{fig:balch}a shows the balance function (Eq. 1) measured for
charged particles for the central and peripheral Au+Au collision
samples. The errors shown are statistical. The systematic error in the
balance function is estimated to be 5\% due to systematic variations in
tracking efficiency, the measurement of pseudorapidity, and
contamination of electrons. The comparable balance function derived for
HIJING events simulated in the STAR detector using GEANT is shown in
Figure \ref{fig:balch}b.

The area under the balance function is constrained by global charge
conservation and STAR's acceptance \cite{Pratt}.  Physical effects over
and above this constraint can be discerned by comparison to a reference
data set that preserves global charge conservation, while removing
effects of dynamical particle correlations.  A relevant reference is
provided in Figure 1b for central and peripheral collision samples
independently by calculating the balance function after the
pseudorapidities of all charged particles within each measured event
have been randomly shuffled.  Dynamical correlations in Au+Au are
reflected in the deviation of the results in Figure 1a from the shuffled
pseudorapidity results in Figure 1b.  In addition Figure 1a also shows
the balance functions generated from conventional mixed-event samples
constructed \cite{marguerite} by choosing random particles from
different measured events with similar event vertex positions and
centralities. The balance function for mixed events integrates to zero
because global charge conservation has been removed.  The fact that the
balance function is zero for all $\Delta \eta$ for all centralities
demonstrates that STAR's acceptance in $\Delta \eta$ is smooth. For both
the mixed events and shuffled pseudorapidity samples, the measured
inclusive pseudorapidity distributions are preserved.

\begin{figure}
\includegraphics[width=3in]{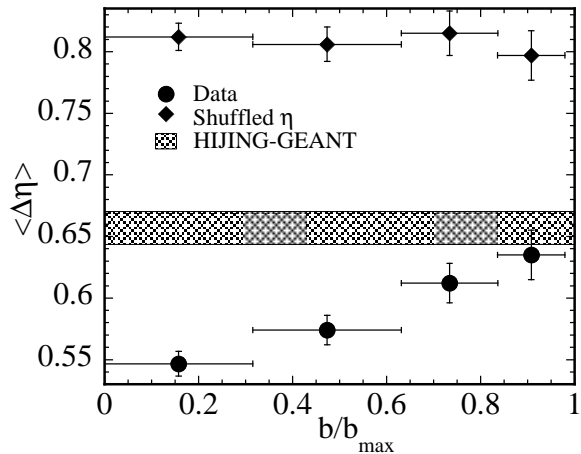}
\caption{\label{fig:balch_summary} The width of the balance function for
charged particles, $\langle \Delta \eta \rangle$, as a function of
normalized impact parameter ($b/b_{max}$).  Error bars shown are
statistical. The width of the balance function from HIJING events is
shown as a band whose height reflects the statistical uncertainty.  Also
shown are the widths from the shuffled pseudorapidity events. }
\end{figure}
Within this area constraint, the variation of the balance function with
centrality can be effectively characterized by the single parameter
$\langle \Delta \eta \rangle$, the mean pseudorapidity difference
weighted by the balance function (excluding the lowest bin in $\Delta
\eta$ to reduce the background correlation from electron contamination).
 We refer to $\langle \Delta \eta \rangle$ below as the ``width" of the
balance function.  The measured widths for four centrality classes are
shown in Figure \ref{fig:balch_summary} as a function of the impact
parameter fraction $b/b_{max}$, which is determined using a simple
geometrical picture \cite{fireball} to relate impact parameter to
fractions of the total cross section. In Figure \ref{fig:balch_summary},
the width of the balance function  measured for central collisions is
significantly smaller than that for peripheral collisions. The results
for the mid-peripheral and mid-central centrality classes decrease
smoothly and monotonically from the peripheral collision value.  Figure
\ref{fig:balch_summary} indicates that while the width observed in
peripheral collisions is consistent with the HIJING prediction, the
balance function for central collisions is significantly narrower,
suggesting a variation in the underlying particle production dynamics
between these two classes of events. In Figure 2 the widths from the
shuffled pseudorapidity events are also shown. These widths show little
centrality dependence and are wider than those of the data or HIJING. 
The widths from shuffled pseudorapidity events represent the maximum
possible width of a balance function measured with the STAR detector.
\begin{figure}
\includegraphics[width=3in]{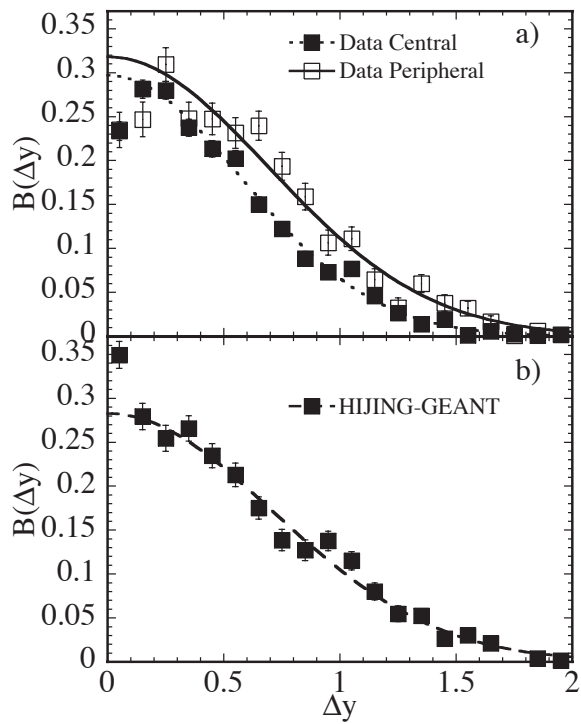}
\caption{\label{fig:balpi}
The balance function versus $\Delta y$ for
identified pion pairs from a) central and peripheral Au+Au collisions at
$\sqrt{s_{NN}}$ = 130 GeV, and b) HIJING events simulated in GEANT
\cite{geant}. To guide the eye, Gaussian fits excluding the lowest two
bins in $\Delta y$ are shown. The error bars shown are statistical. The
balance function for HIJING events is independent of centrality.}
\end{figure}

The results for identified charged pion pairs are similar to those for
non-identified charged particles as indicated in Figure \ref{fig:balpi}.
The overall shape of the balance function is similar to that in Figure
\ref{fig:balch}. However, the data for pions have a dip near $\Delta y =
0$ (Figure \ref{fig:balpi}a). As shown in reference \cite{Pratt}, this
dip can be understood as the combined effect of Bose-Einstein
correlations and Coulomb interactions between charged pions. HIJING does
not account for these effects, and the balance function predicted in
Figure \ref{fig:balpi}b therefore does not show a dip, although the
enhancement of the lowest bin due to electron contamination is still
apparent.

\begin{figure}
\includegraphics[width=3in]{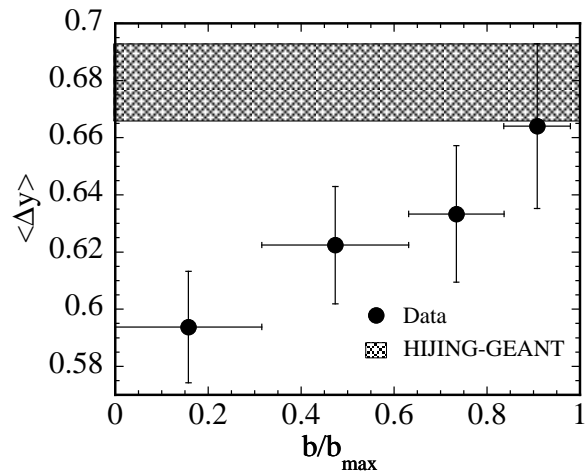}
\caption{\label{fig:balpi_summary}
The width of the balance function for
identified charged pions, $\langle \Delta y \rangle$, as a function of
normalized impact parameter ($b/b_{max}$).  Error bars shown are
statistical. The width of the balance function from HIJING events is
shown as a band whose height reflects the statistical uncertainty.}
\end{figure}

The width of the balance functions for the four centrality classes is
shown in Figure \ref{fig:balpi_summary}, where the lowest two bins in
$\Delta y$ have been excluded to avoid the effects of Bose-Einstein
correlations and Coulomb interactions and enable a valid comparison with
HIJING. The results indicate that the width of the balance function for
peripheral events is consistent with that expected from HIJING, while
the width for central events is significantly smaller as was observed
with charged particle pairs.

To gain some insight into the observed widths in central collisions, we
compared with the thermal model (final temperature of 165 MeV) presented
in \cite{balance_paper} filtered through the acceptance of STAR.  We
find that the predicted widths are larger than those we observe in
central collisions. Thermal model calculations have also been done at
low final temperature (105 MeV) and high transverse velocity (0.77 c),
while maintaining the same average transverse momentum. These
calculations predict a balance function width consistent with our
observations in central collisions.
Thus the observed narrowing of the
balance function in central collisions can be parametrized in a thermal
model incorporating both strong transverse flow and a constraint  that
the balancing particles are emitted close together in space-time
\cite{Pratt_workshop}. This constraint could arise from delayed
hadronization compared with the characteristic 1 fm/c hadronization time
or from some other phenomena such as anomalously short diffusion of
particle pairs \cite{Pratt_workshop}.  Other effects besides flow may
also narrow the balance function with centrality.  One effect is the
decay of resonances such as the $\rho ^{0}$ that have lifetimes similar
to the proposed time of delayed hadronization.  Another effect could be
opacity \cite{opacity} where pairs with large opening angles are
absorbed leading to a narrower balance function.

In summary, measurement of the balance function for Au+Au collisions at
$\sqrt{s_{NN}}$=130 GeV has been stimulated by the prediction
\cite{balance_paper} that the width of the balance function should be
significantly reduced by late hadronization.  We indeed observe a
narrowing of the balance function for more central collisions for all
charged particle and for charged pion pairs.  Only for peripheral
collisions is the width consistent with HIJING predictions treating the
Au+Au collision as a superposition of independent nucleon-nucleon
scatterings.  Interpretation of the observed narrowing requires more
detailed study of its sensitivity to such other effects as flow,
resonance production and opacity, in addition to late
hadronization.

\begin{acknowledgments}
We wish to thank the RHIC Operations Group and the RHIC Computing
Facility
at Brookhaven National Laboratory, and the National Energy
Research 
Scientific Computing Center at Lawrence Berkeley National
Laboratory
for their support. This work was supported by the Division of
Nuclear 
Physics and the Division of High Energy Physics of the Office
of Science of 
the U.S. Department of Energy, the U.S. National Science
Foundation,
the Bundesministerium fuer Bildung und Forschung of
Germany,
the Institut National de la Physique Nucleaire et de la
Physique 
des Particules of France, the United Kingdom Engineering and
Physical 
Sciences Research Council, Fundacao de Amparo a Pesquisa do
Estado de Sao 
Paulo, Brazil, the Russian Ministry of Science and
Technology, the
Ministry of Education of China, the National Natural
Science Foundation 
of China, and the Swedish National Science
Foundation.
\end{acknowledgments}

\end{document}